\documentclass{ws-ijmpd}

\begin{document}

\markboth{\.{I}. Yavuz, \.{I}. Y{\i}lmaz and H. Baysal} {Strange
Quark Matter Attached to the String Cloud In the Spherical
Space-time \ldots}

%
\catchline{}{}{}{}{}

\title{STRANGE QUARK MATTER ATTACHED TO THE STRING CLOUD IN THE SPHERICAL
SYMMETRIC SPACE-TIME ADMITTING CONFORMAL MOTION}

\author{\.{I}LHAM\.{I} YAVUZ}

\address{Maltepe University, Engineering and Architecture Faculty, Department of Computer Engineering,
Maltepe-\.Istanbul, Turkey\\
 iyavuz@maltepe.edu.tr}

\author{\.{I}HSAN YILMAZ\dag\S$\;$ and H\"{U}SN\"{U} BAYSAL\ddag\S}
\address{\dag\c{C}anakkale Onsekiz Mart University, Arts and Sciences
Faculty, Department of Physics, Terzio\u{g}lu Campus, 17100
\c{C}anakkale, Turkey\\
iyilmaz@comu.edu.tr\\
\ddag\c{C}anakkale Onsekiz Mart University, Arts and Sciences
Faculty, Department of Mathematics, Terzio\u{g}lu Campus, 17100
\c{C}anakkale, Turkey\\
hbaysal@comu.edu.tr\\
\S\c{C}anakkale Onsekiz Mart University, Astrophysics Research
Center, 17100 \c{C}anakkale,
Turkey\\
}

\maketitle

\begin{history}
\received{Day Month Year} \revised{Day Month Year}
\end{history}

\begin{abstract}
In this paper, we have examined charged strange quark matter
attached to the string cloud in the spherical symmetric space-time
admitting one-parameter group of conformal motions. For this
purpose, we have solved Einstein's field equations for spherical
symmetric space-time with strange quark matter attached to the
string cloud via conformal motions. Also, we have discussed the
features of the obtained solutions.
\end{abstract}

\keywords{Strange quark matter; string; conformal motion.}

\section{Introduction}

It is still a challenging problem to know the exact physical
situation at very early stages of the formation of our universe. At
the very early stages of evolution of universe, it is generally
assumed that during the phase transition (as the universe passes
through its critical temperature) the symmetry of the universe is
broken spontaneously. It can give rise to topologically stable
defects such as domain walls, strings and
monopoles.\cite{Kibble}$^{,}$\cite{Dolgov}

Of all these cosmological structures, cosmic strings have excited
the most interest. The present day configurations of the universe
are not contradicted by the large scale network of strings in the
early universe. Moreover, they may act as gravitational lenses and
may give rise to density fluctuations leading to the formations of
galaxies. \cite{Vilenkin} Strings posses stress energy and are
coupled to the gravitational field.

In String Theory, the myriad of particle types is replaced by a
single fundamental building block, a 'string'. These strings can be
closed, like loops, or open, like a hair. As the string moves
through time it traces out a tube or a sheet, according to whether
it is closed or open, Furthermore, the string is free to vibrate,
and different vibrational modes of the string represent the
different particle types, since different modes are seen as
different masses or spins.

One mode of vibration, or 'note', makes the string appear as an
electron, another as a photon. There is even a mode describing the
graviton, the particle carrying the force of gravity,  which is an
important reason why string theory has received so much attention.
The point is that we can make sense of the interaction of two
gravitons in string theory in a way we could not in QFT. There are
no infinities! And gravity is not something we put in by hand. It
has to be there in a theory of strings. So, the first great
achievement of string theory was to give a consistent theory of
quantum gravity, which resembles GR at macroscopic distances.
Moreover string theory also possesses the necessary degrees of
freedom to describe the other interactions! At this point a great
hope was created that string theory would be able to unify all the
known forces and particles together into a single 'Theory of
Everything'.

In this study, we will attach strange quark matter to the string
cloud. It is plausible to attach strange quark matter to the string
cloud. Because, one of such transitions during the phase transitions
of the universe could be Quark Gluon Plasma
$(QGP)\rightarrow\textrm{hadron gas}$ (called quark-hadron phase
transition) when cosmic temperature was $T\sim200$ MeV.

Recently, Masaharu and Fukutome\cite{Masaharu} and Narodetski
\textit{et al.}\cite{Narodetski} have studied tetra quark particle
and pentaquarks in the string model, respectively. Also, Harko and
Cheng have studied strange quark matter adopting in the form of
perfect fluid in the spherical symmetric space-times.\cite{Harko}

The possibility of the existence of quark matter dates back to early
seventies. Bodmer\cite{Bodmer} and Witten\cite{Witten} have proposed
two ways of formation of strange matter: the quark-hadron phase
transition in the early universe and conversion of neutron stars
into strange ones at ultrahigh densities. In the theories of strong
interaction quark bag models suppose that breaking of physical
vacuum takes place inside hadrons. As a result vacuum energy
densities inside and outside a hadron become essentially different,
and the vacuum pressure on the bag wall equilibrates the pressure of
quarks, thus stabilizing the system. If the hypothesis of the quark
matter is true, then some of neutrons stars could actually be
strange stars, built entirely of strange
matter.\cite{Alcock}$^{,}$\cite{Haensel}

Typically, strange quark matter is modeled with an equation of state
(EOS) based on the phenomenological bag model of quark matter, in
which quark confinement is described by an energy term proportional
to the volume.

In this model, quarks are though as degenerate Fermi gases, which
exist only in a region of space endowed with a vacuum energy density
$B_{c}$ (called as the bag constant). Also, in the framework of this
model the quark matter is composed of massless u, d quarks, massive
s quarks and electrons.

In the simplified version of the bag model, assuming quarks are
massless and non interacting, we then have quark pressure
$p_{q}=\rho_{q}/3$ ($\rho_{q}$ is the quark energy density); the
total energy density is $\rho=\rho_{q}+B_{c}$ but total pressure is
$p=p_{q}-B_{c}$.

In this paper, we will solve Einstein's field equations for
spherical symmetric space-times with strange quark matter attached
to the string cloud via Conformal Killing Vector (CKV).

General Relativity provides a rich arena to use symmetries in order
to understand the natural relation between geometry and matter
furnished by the Einstein equations. Symmetries of
geometrical/physical relevant quantities of this theory are known as
collineations. The most useful collineations is conformal killing
vectors. So, in this paper it is imposed the condition that the
space-time manifold admits a conformal killing vector. Conformal
Killing Vectors provide a deeper insight into the space-time
geometry and facilitate generation of exact solutions to the field
equations.

Conformal collineation is defined by,
\begin{equation}
\pounds_{\xi}g_{ab}=2\psi g_{ab} \quad \textrm{,}\quad
\psi=\psi(x^{a}),\label{eq1}
\end{equation}
where $\pounds_{\xi}$ signifies the Lie derivative along $\xi^{a}$
and $\psi(x^{a})$ is the conformal factor. In particular, $\xi$ is a
special conformal Killing vector (SCKV) if $\psi_{;ab}=0$ and
$\psi_{,a}\neq0$. Other subcases are homothetic vector (HV) if
$\psi_{,a}=0$ and $\psi\neq0$, and Killing vector (KV) if $\psi=0$.
Here ; and , denote the covariant and ordinary derivatives,
respectively.

The paper is outlined as follows. In section 2, Einstein field
equations are obtained for charged strange quark matter attached to
the string cloud in the spherical symmetric space-time. In section 3
Einstein field equations are solved for the same matter via
conformal motions depending on conformal factor i.e., $\psi(x^{a})$.
In section 4, concluding remarks are given.

\section{Einstein's Field Equations}

Let us consider a static distribution of matter represented by
charged spherical symmetric matter which may be strange quark matter
attached to the string cloud.

In Schwarzschild coordinates the line element takes the following
form:
\begin{equation}
ds^{2}=e^{\nu(r)}dt^{2}-e^{\lambda(r)}dr^{2}-r^{2}d\Omega^{2},
\label{eq2}
\end{equation}
with
\[d\Omega^{2}=d\theta^{2}+\sin^{2} \theta d\phi^{2}, \qquad
 x^{1,2,3,4}\equiv r,\theta,\phi,t \]
The total energy-momentum tensor $T_{ab}$ is assumed to be the sum
of two parts, $T_{ab}^{S}$ and $T_{ab}^{E}$, for string cloud and
electromagnetic contributions, respectively, i.e.,
\begin{equation}
T_{ab}=T_{ab}^{S}+T_{ab}^{E}. \label{eq3}
\end{equation}
The energy-momentum tensor for string cloud \cite{Letelier} is given
by
\begin{equation}
T_{ab}^{S}=\rho U_{a}U_{b}-\rho_{s}X_{a}X_{b} \label{eq4}
\end{equation}
here $\rho$ is the rest energy for the cloud of strings with
particles attached to them and $\rho_{s}$ is string tension density;
they are related by
\begin{equation}
\rho=\rho_{p}+\rho_{s} \quad \textrm{or}\quad \rho_{p}=\rho-\rho_{s}
\label{eq5}
\end{equation}
where $\rho_{p}$ is the particle energy density.

In this paper we will take strange quark matter energy density
instead of particle energy density in the string cloud. In this case
from Eq. (\ref{eq5}), we get
\begin{equation}
\rho=\rho_{q}+\rho_{s} +B_{c}\quad \textrm{or}\quad
\rho_{q}+B_{c}=\rho - \rho_{s} \label{eq6}
\end{equation}

If we put Eq. (\ref{eq6}) into Eq. (\ref{eq4}), we have for strange
quark matter attached to the string cloud
\begin{equation}
T_{ab}^{S}= (\rho_{q}+\rho_{s}+B_{c})U_{a}U_{b}-\rho_{s} X_{a}X_{b}
\label{eq7}
\end{equation}
where $U^{a}$ is the four velocity $U^{a}=\delta_{4}^{a}e^{-\nu/2}$,
$X^{a}$  is the unit spacelike vector in the radial direction
$X^{a}=\delta_{1}^{a}e^{-\lambda/2}$ which represent the strings
directions in the cloud,i.e. the direction of anisotropy.
\begin{equation}
T_{ab}^{E}=- \frac{1}{4\pi}
\left(F_{a}^{c}F_{bc}-\frac{1}{4}g_{ab}F_{ef}F^{ef} \right)
\label{eq8}
\end{equation}
where $F_{ab}$ is the electromagnetic field tensor defined in terms
of the four-potential $A_{a}$ as
\[
F_{ab}=A_{b;a}-A_{a;b}.
\]

For the electromagnetic field we shall adopt the gauge
\[
A_{a}(0,0,0,\phi(r)).
\]

Einstein-Maxwell equations can be expressed as
\begin{eqnarray}
& & \ R_{ab}-\frac{1}{2}Rg_{ab}= 8\pi T_{ab},\label{eq9} \\
& & \ F_{ab;c}+F_{bc;a}+F_{ca;b}=0,\label{eq10} \\
& & \ F_{;b}^{ab}=-4\pi J^{a},\label{eq11}
\end{eqnarray}
where $J^{a}$ is the four-current density that becomes
$J^{a}=\overline{\rho}_{e}U^{a}$ and $\overline{\rho}_{e}$ is the
proper charge density.

Using the line element (\ref{eq2}), the field equations
(\ref{eq7})-(\ref{eq11}) take the form,
\begin{eqnarray}
& & \ 8\pi\rho+E^{2}=-e^{-\lambda} \left(\frac{1}{r^{2}}-\frac{\lambda^{'}}{r} \right)+\frac{1}{r^{2}}, \label{eq12} \\
& & \ -8\pi\rho_{s}+E^{2}=-e^{-\lambda} \left(\frac{1}{r^{2}}+\frac{\nu^{'}}{r} \right)+\frac{1}{r^{2}}, \label{eq13} \\
& & \ E^{2}=\frac{e^{-\lambda}}{2} \left(\nu^{''}+\frac{\nu^{'2}}{2}+\frac{\nu^{'}-\lambda^{'}}{r}-\frac{\nu^{'}\lambda^{'}}{2} \right),\label{eq14} \\
& & \ \left[r^{2}E(r)\right]'=4\pi \rho_{e}r^{2}.\label{eq15}
\end{eqnarray}
Primes denote differentiation with respect to $r$, and $E$ is the
usual electric field intensity defined as
\begin{eqnarray}
& & \ F_{41}F^{41}=-E^{2},\nonumber \\
& & \ E(r)=-e^{-(\nu+\lambda)/2}\phi^{'}(r),\label{eq16} \\
& & \ \phi^{'}(r)=F_{14}=-F_{41}.\nonumber
\end{eqnarray}
The charge density $\rho_{e}$ defined in Eq. (\ref{eq15}) is related
to the proper charge density $\overline{\rho}_{e}$ by
\begin{equation}\label{eq17}
\rho_{e}=\overline{\rho}_{e}e^{\lambda/2}.
\end{equation}

\section{Solutions of the Field Equations}

Now we shall assume that space-time admits a one-parameter group of
conformal motions (Eq. (\ref{eq1})), i.e.,
\begin{equation}
\pounds_{\xi}g_{ab}=\xi_{a;b}+\xi_{b;a}=\psi g_{ab},\label{eq18}
\end{equation}
where $\psi$ is an arbitrary functions of $r$. From Eqs. (\ref{eq2})
and (\ref{eq18}) and by virtue of spherical symmetry, we get the
following expressions
\begin{eqnarray}
& & \ \xi^{1}\nu'=\psi,\label{eq19} \\
& & \ \xi^{4}=C_{1}=const,\label{eq20}\\
& & \ \xi^{1}=\psi r/2,\label{eq21} \\
& & \ \lambda'\xi^{1}+2\xi^{1} _{,1}=\psi,\label{eq22}
\end{eqnarray}
where a comma denotes partial derivatives. From Eqs.
(\ref{eq19})-(\ref{eq22}), we get
\begin{eqnarray}
& & e^{\nu}=C_{2}^{2}r^{2},\label{eq23} \\
& & e^{\lambda}=\left(\frac{C_{3}}{\psi}\right)^2,\label{eq24}\\
& & \xi^{a}=C_{1}\delta_{4}^{a}+(\psi
r/2)\delta_{1}^{a},\label{eq25}
\end{eqnarray}
where $C_{2}$ and $C_{3}$ are constants of
integration.\cite{Herrera} Expressions (\ref{eq23})-(\ref{eq25})
contain all the implications derived from the existence of the
conformal collineation.

Now substituting (\ref{eq23}) and (\ref{eq24}) into Eqs.
(\ref{eq12})-(\ref{eq14}), we have
\begin{eqnarray}
& & \rho+E^{2}=(1/r^{2})(1-\psi^{2}/C_{3}^{2})-2\psi\psi'/C_{3}^{2}r,\label{eq26} \\
& & \rho_{s}+E^{2}=(1/r^{2})(1-3\psi^{2}/C_{3}^{2}),\label{eq27}\\
& & E^{2}=\psi^{2}/C_{3}^{2}r^{2}+2\psi\psi'/C_{3}^{2}r.\label{eq28}
\end{eqnarray}
Here we use geometrized unit so that ${8\pi G=c=1}$.  From Eqs.
(\ref{eq6}) and (\ref{eq26})-(\ref{eq28}) we get
\begin{eqnarray}
& & \rho=\frac{1}{r^2}\left(1-\frac{2\psi^{2}}{C_{3}^{2}}\right)-\frac{4\psi\psi'}{C_{3}^{2}r},\label{eq29} \\
& & \rho_{s}=\frac{1}{r^2}\left(1-\frac{4\psi^{2}}{C_{3}^{2}}\right)-\frac{2\psi\psi'}{C_{3}^{2}r},\label{eq30}\\
& & E^{2}=\frac{\psi^{2}}{C_{3}^{2}r^{2}}+\frac{2\psi\psi'}{C_{3}^{2}r},\label{eq31}\\
& &
\rho_{p}=\rho_{q}+B_{c}=\rho-\rho_{s}=\frac{2}{r}\left(\frac{\psi^{2}}{C_{3}^{2}}-\frac{\psi\psi'}{C_{3}^{2}}\right).\label{eq32}
\end{eqnarray}

Using Eqs. (\ref{eq23}) and (\ref{eq24}), the line element (Eq.
(\ref{eq2})) becomes
\begin{equation}
ds^{2}=C_{2}^{2}r^{2}dt^{2}-\frac{C_{3}^{2}}{\psi^{2}}dr^{2}-r^{2}d\Omega^{2},\label{eq33}
\end{equation}

If the function $\psi$ and an equation of state for the stresses are
specified a priori, the problem will be fully determined. So, We
will examine the following physically meaningful three cases
depending on $\psi(r)$.

{\bf Case ({\it i})} If $\psi=C_{4}r$ then from Eqs.
(\ref{eq29})-(\ref{eq32}) we get
\begin{eqnarray}
& & \rho=\rho_{s}=\frac{1}{r^{2}}-6 \left(\frac{C_{4}}{C_{3}} \right)^{2}, \label{eq34}\\
& & E^{2}=3 \left(\frac{C_{4}}{C_{3}} \right)^{2} \quad
\textrm{and}\quad \rho_{p}=0.\label{eq35}
\end{eqnarray}
where $C_{4}$ is integration constant.

Let us now consider that the charged sphere extends to radius
$r_{0}$. Then the solution of Einstein-Maxwell equations for
$r>r_{0}$ is given by the Reissner-Nordstr\"{o}m metric as
\begin{equation}
ds^{2}=\left(1-\frac{2M}{r}+\frac{q^{2}}{r^{2}}\right)dt^{2}-\left(1-\frac{2M}{r}+\frac{q^{2}}{r^{2}}\right)^{-1}dr^{2}-r^{2}d\Omega^{2},\label{eq36}
\end{equation}
and the radial electric field is
\begin{equation}
E=q/r^{2},\label{eq37}
\end{equation}
where $M$ and $q$ are the total mass and charge, respectively.

To match the line element (\ref{eq33}) with the
Reissner-Nordstr\"{o}m metric across the boundary $r=r_{0}$ we
require continuity of gravitational potential $g_{ab}$ at $r=r_{0}$

\begin{equation}
(C_{2}r_{0})^{2}=\left(\frac{\psi}{C_{3}}\right)^{2}=1-\frac{2M}{r_{0}}+\frac{q^{2}}{r_{0}^{2}},\label{eq38}
\end{equation}
and also we require the continuity of the electric field, which
leads to
\begin{equation}
E(r_{0})=\frac{q}{r_{0}^{2}},\label{eq39}
\end{equation}

From Eq. (\ref{eq39}) and left hand side of Eq. (\ref{eq35}) we get
\begin{equation}
\frac{q^{2}}{r_{0}^{2}}=3r_{0}^{2}\left(\frac{C_{4}}{C_{3}}\right)^{2},\label{eq40}
\end{equation}

Feeding this expression back into Eq. (\ref{eq38}) we obtain
\begin{equation}
\frac{M}{r_{0}}=\frac{1}{2}+\left(\frac{C_{4}}{C_{3}}\right)^{2}r_{0}^{2},\label{eq41}
\end{equation}
or from Eqs. (\ref{eq40}) and (\ref{eq41}) we have
\begin{equation}
M=\frac{r_{0}}{2}+\frac{1}{3}\frac{q^{2}}{r_{0}},\label{eq42}
\end{equation}

{\bf Case ({\it ii})} If
$\psi=\frac{1}{2}\sqrt{C_{3}^{2}+\frac{4C_{5}}{r^{4}}}$, from Eqs.
(\ref{eq29})-(\ref{eq32}) we get the following expressions
\begin{eqnarray}
& &
\rho=\rho_{p}=\frac{1}{2r^{2}}+\frac{6C_{5}}{C_{3}^{2}r^{6}}\label{eq43}\\
& & \textrm{or} \nonumber\\
& & \rho_{q}=\frac{1}{2r^{2}}+\frac{6C_{5}}{C_{3}^{2}r^{6}}-B_{c},\label{eq44}\\
& & E^{2}=\frac{1}{4r^{2}}-\frac{3C_{5}}{C_{3}^{2}r^{6}} \quad
\textrm{and}\quad \rho_{s}=0.\label{eq45}
\end{eqnarray}
where $C_{5}$ is another integration constant.

Setting $\frac{C_{5}}{C_{3}^{2}}=\alpha r_{0}^{4}$ and using Eqs.
(\ref{eq37}), (\ref{eq38}) and (\ref{eq45}) we obtain the total mass
and the total charge:
\begin{eqnarray}
& & \frac{q^{2}}{r_{0}^{2}}=\frac{(1-12\alpha)}{4} \quad ,\quad 0\leq\alpha<\frac{1}{12} \nonumber \\
& & \frac{M}{r_{0}}=\frac{1}{2}-2\alpha \nonumber
\end{eqnarray}

{\bf Case ({\it iii})} If $\psi=\frac{C_{6}}{\sqrt{r}}$, from Eqs.
(\ref{eq29})-(\ref{eq32}) we get the following expressions,
\begin{eqnarray}
& & \rho=\frac{1}{r^{2}},\label{eq46} \\
& & \rho_{s}=\frac{1}{r^{2}}-\frac{3C_{6}^{2}}{C_{3}^{2}r^{3}},\label{eq47} \\
& & \rho_{p}=\frac{3C_{6}^{2}}{C_{3}^{2}r^{3}} \quad
\textrm{or}\quad \rho_{q}=\frac{3C_{6}^{2}}{C_{3}^{2}r^{3}}-B_{c}
\quad \textrm{and}\quad E^{2}=0 \label{eq48}
\end{eqnarray}
where $C_{6}$ is a constant.

\section{Concluding Remarks}
In this paper, we have studied charged strange quark matter attached
to the string cloud in the spherical symmetric space-time admitting
one-parameter group of conformal motions.

We have obtained the following properties.
\begin{romanlist}
\item [(a)] $e^{\mu}$ and $e^{\lambda}$ are positive, continuous and
nonsingular for $r<r_{0}$.

\item [(b)] In the case (i) and case (ii) we have matched our solutions
with the Reissner-Nordstr\"{o}m metric at $r=r_{0}$. In the case (i)
we have had charged geometric string solutions and charged black
string solutions (see Eq. (\ref{eq34}) and (\ref{eq42}) ). In this
case, we have obtained the increase of the total mass caused by the
charge (see Eq. (\ref{eq42})). Also, if $q=0$ we get total mass for
noncharged black string, i.e. Schwarzschild like black string.

\item [(c)] In the case (ii) and case (iii) we have obtained pressureless
charged strange quark matter solution and non charged strange quark
matter attached to the string cloud, respectively.

\item [(d)] In the case (ii), the requirement $\rho_{p}>0$ throughout the
distribution implies that $C_{5}$ must be non-negative. However, Eq.
(\ref{eq45}) gives $E^{2}<0$ in the central region. Thus we have
obtained an analytical solution of Einstein-Maxwell equations in the
region $r^{4}>12 C_{5}/C_{3}^{2}$.

\item [(e)] In the case (iii) from Eqs. (\ref{eq46})-(\ref{eq48}) we may
conclude that strange quark matter decreases the energy of the
string.
\end{romanlist}

\end{document}